\documentclass[twocolumn]{aastex63}
\usepackage{enumitem}
\usepackage{amsmath}
\usepackage[caption=false]{subfig}

\received{*}
\revised{*}
\accepted{*}
\shorttitle{Morphology of Obscured AGN}
\shortauthors{Lambrides et al.}
\graphicspath{{./}{figures/}}

\begin{document}

\title{Lower-Luminosity Obscured AGN Host Galaxies are Not Predominantly in Major-Merging Systems at Cosmic Noon}

\correspondingauthor{Erini Lambrides}
\email{erini.lambrides@jhu.edu}

\author[0000-0002-0786-7307]{Erini L. Lambrides}
\affil{Department of Physics and Astronomy, Johns Hopkins University, Bloomberg Center, 3400 N. Charles St., Baltimore, MD 21218, USA}

\author{Marco Chiaberge}
\affil{AURA for the European Space Agency (ESA), ESA Office, Space Telescope Science Institute, 3700 San Martin Drive, Baltimore, MD 21218, USA}
\affil{Department of Physics and Astronomy, Johns Hopkins University, Bloomberg Center, 3400 N. Charles St., Baltimore, MD 21218, USA}

\author{Timothy Heckman}
\affil{Department of Physics and Astronomy, Johns Hopkins University, Bloomberg Center, 3400 N. Charles St., Baltimore, MD 21218, USA}

\author{Allison Kirkpatrick}
\affil{Department of Physics and Astronomy, University of Kansas, Lawrence, KS 66045, USA}

\author{Eileen T. Meyer}
\affil{Department of Physics, University of Maryland, Baltimore County, 1000 Hilltop Circle, Baltimore, MD 21250, USA}

\author{Andreea Petric}
\affil{Space Telescope Science Institute, 3700 San Martin Drive, Baltimore, MD 21218, USA}

\author{Kirsten Hall}
\affil{Schmidt Science Fellow}
\affil{Atomic and Molecular Physics Division, Harvard-Smithsonian Center for Astrophysics, 60 Garden Street, Cambridge, MA, 02138, USA}

\author{Arianna Long}
\affil{Department of Physics and Astronomy, University of California, Irvine, CA 92697, USA}

\author[0000-0002-5437-6121]{Duncan J. Watts}
\affil{Institute of Theoretical Astrophysics, University of Oslo, P.O. Box 1029 Blindern, N-0315 Oslo, Norway}

\author{Roberto Gilli}
\affil{INAF – Osservatorio di Astrofisica e Scienza dello Spazio di Bologna, Via P. Gobetti 93/3, 40129 Bologna, Italy}

\author{Raymond Simons}
\affil{Space Telescope Science Institute, 3700 San Martin Drive, Baltimore, MD 21218, USA}

\author{Kirill Tchernyshyov}
\affil{Department of Astronomy, University of Washington, Seattle, WA, USA}

\author{Vicente Rodriguez-Gomez}
\affil{Instituto de Radioastronom\'ia y Astrof\'isica, Universidad Nacional Aut\'onoma de M\'exico, Apdo. Postal 72-3, 58089 Morelia, Mexico}

\author{Fabio Vito}
\affil{Scuola Normale Superiore, Piazza dei Cavalieri 7, 56126, Pisa, Italy}

\author{Alexander de la Vega}
\affil{Department of Physics and Astronomy, Johns Hopkins University, Bloomberg Center, 3400 N. Charles St., Baltimore, MD 21218, USA}

\author{Jeffrey R. Davis}
\affil{Department of Physics and Astronomy, Johns Hopkins University, Bloomberg Center, 3400 N. Charles St., Baltimore, MD 21218, USA}

\author{Dale D Kocevski}
\affil{Department of Physics and Astronomy, Colby College, 5800 Mayflower Hill, Waterville, ME 04961, USA}

\author{Colin Norman}
\affil{Space Telescope Science Institute, 3700 San Martin Drive, Baltimore, MD 21218, USA}
\affil{Department of Physics and Astronomy, Johns Hopkins University, Bloomberg Center, 3400 N. Charles St., Baltimore, MD 21218, USA}

\begin{abstract}

For over 60 years, the scientific community has studied actively growing central super-massive black holes (active galactic nuclei -- AGN) but fundamental questions on their genesis remain unanswered. Numerical simulations and theoretical arguments show that black hole growth occurs during short-lived periods ($\sim$ 10$^{7}$ -10$^{8}$ yr) of powerful accretion. Major mergers are commonly invoked as the most likely dissipative process to trigger the rapid fueling of AGN. If the AGN-merger paradigm is true, we expect galaxy mergers to coincide with black hole accretion during a heavily obscured AGN phase (N$_H$ $ > 10^{23}$ cm$^{-2}$). Starting from one of the largest samples of obscured AGN at 0.5 $<$ $z$ $<$ 3.1, we select 40 non-starbursting lower-luminosity obscured AGN. We then construct a one-to-one matched redshift- and near-IR magnitude- matched non-starbursting inactive galaxy control sample. Combining deep color \textit{Hubble Space Telescope} imaging and a novel method of human classification, we test the merger-AGN paradigm prediction that heavily obscured AGN are strongly associated with galaxies undergoing a major merger. On the total sample of 80 galaxies, we estimate each individual classifier's accuracy at identifying merging galaxies/post-merging systems and isolated galaxies. We calculate the probability of each galaxy being in either a major merger or isolated system, given the accuracy of the human classifiers and the individual classifications of each galaxy. We do not find statistically significant evidence that obscured AGN at cosmic noon are predominately found in systems with evidence of significant merging/post-merging features.   

\end{abstract}


\section{Introduction} \label{sec:intro}

Super-massive blackholes (SMBHs) are essentially in every massive galaxy, and when they are actively accreting matter, known as active galactic nuclei (AGN), they can potentially inject energy into the gas and expel it and/or prevent it from cooling and collapsing into stars\citep[e.g.][]{bower06,croton,heckmanaa}. Matter must lose almost all ($\sim$99.9\%) of its angular momentum in order to accrete onto the SMBH, thus, studying dissipative processes such as mergers, tidal interactions, stellar bars and disk instabilities is central to understanding the details of AGN fueling. Despite distinct differences between dissipative processes, neither observational nor theoretical studies converge on a dominant mechanism for funneling matter onto the central SMBH \citep{jogee}. Galaxy mergers with comparable mass ratios ($\geq 1:4$ also defined as major mergers) are one of the most popular mechanisms invoked, yet the observational consensus is mixed. While some empirical and theoretical studies find a connection between mergers and ultra-luminous infrared galaxies \citep{sanders96, veilleux09,lacy2018}, local AGN \citep{koss10,ellison13,ellison19,gao2020}, high-luminosity AGN \citep{urrutia08,treister12,glikman15,donley18}, and radio-loud AGN \citep{chiaberge15}, others find no connection between mergers and X-ray detected AGN \citep{gabor09,georgakakis09}, high-luminosity AGN \citep{villforth14,villforth17,marian19}, and low-to intermediate luminosity AGN (L$_{\textrm{Bol}} < 10^{44}$ ergs/s) \citep{grogin05,schawinski11,rosario15}.  

It is possible that the AGN-merger connection has been systematically missed in some studies due to poor sampling of obscured AGN. If the AGN-merger paradigm is true, then we can expect a heavily obscured accretion AGN phase to coincide with galaxy coalescence \citep{sanders96,cattaneo05,hopkins08}. In other words, if a major-merger triggers most AGN, then AGN behind large neutral hydrogen column densities ($N_\mathrm{H} > 10^{23}$ cm$^{-2}$) should exist in association with the most spectacular phases of mergers. 

Obscured sources are inherently difficult to detect in the X-rays, but through the combination of large and deep X-ray surveys with other multi-wavelength observations, a large obscured AGN sample can be constructed. X-ray observations are thought to provide one of the most reliable methods of both selecting AGN and estimating the amount of AGN obscuration \citep{brandt05,xue11,liu}; however this is not always true, as \cite{comastri11,donley12} show that even some of the deepest X-ray surveys miss a substantial fraction ($\sim 40$\%) of heavily obscured objects. 

One of the first studies of its kind, \cite{kocev15} analyzed a sample of obscured AGN defined using a selection based on X-ray data from the \textit{Chandra X-ray Observatory}, with deepest observations at 4Ms \citep{xue11}. Using a single \textit{Hubble Space Telescope}(HST) near-IR(NIR) band, they found evidence that heavily obscured AGN are more likely to be in mergers than their a less obscured AGN counterparts: point-sources included 21.5\%$^{+4.2\%}_{-3.3\%}$ heavily obscured AGN in mergers versus 7.8\%$^{+1.9\%}_{-1.3\%}$. 

In 2017, the 7MS Chandra Deep-Field South Survey (7MCDFS), the deepest X-ray survey ever conducted was released. Within this substantially deeper catalog and the combination of IR, optical, and radio data-sets, \citet{lambrides20} find that $30\%$ of the X-ray-detected AGN were mis-classified as low-luminosity un-obscured AGN. \citet{lambrides20} argues that these objects instead represent the faintest, and potentially the most obscured AGN in the 7MCDFS sample. It is imperative that we morphologically analyze these objects whose addition may either lend or remove credence to the obscured AGN-merger paradigm. 

In this work we combine the \citet{lambrides20} obscured AGN sample with publicly available HST imaging to determine the merger status of the host galaxies of obscured AGN. The first paper in our series, \citet{lambrides21} (L21) introduced a novel statistical method where the accuracy of human classifiers are taken into account in a Bayesian probabilistic framework to determine the merger fraction and individual probabilities of a galaxy being in a merging system. In section 2 we describe the obscured AGN sample, the control sample, the HST data, and the simulated data used in this work. In section 3, we describe the survey framework and statistical models used to derive a merger fraction of a population. In section 4, we present the results of the merger fraction of the obscured AGN population. In section 5, we discuss how our results compare to other studies and the implications our results have on AGN triggering models. In section 6, we present the summary and conclusion. We use an $h = 0.7$, $\Omega_{m} = 0.3$, $\Omega_{\Lambda} = 0.7$ cosmology throughout this paper. We use the k-sample Anderson-Darling mid-rank statistic to test the null hypothesis that two samples are drawn from the same population, and report the test statistic (D$_{ADK}$) significance level at which the null hypothesis for the provided samples can be rejected \citep{ksamp}.

\section{Sample Selection and Datasets}

\subsection{Heavily Obscured AGN} \label{sec:obs_sample}


\begin{figure*}
\centering
\includegraphics[]{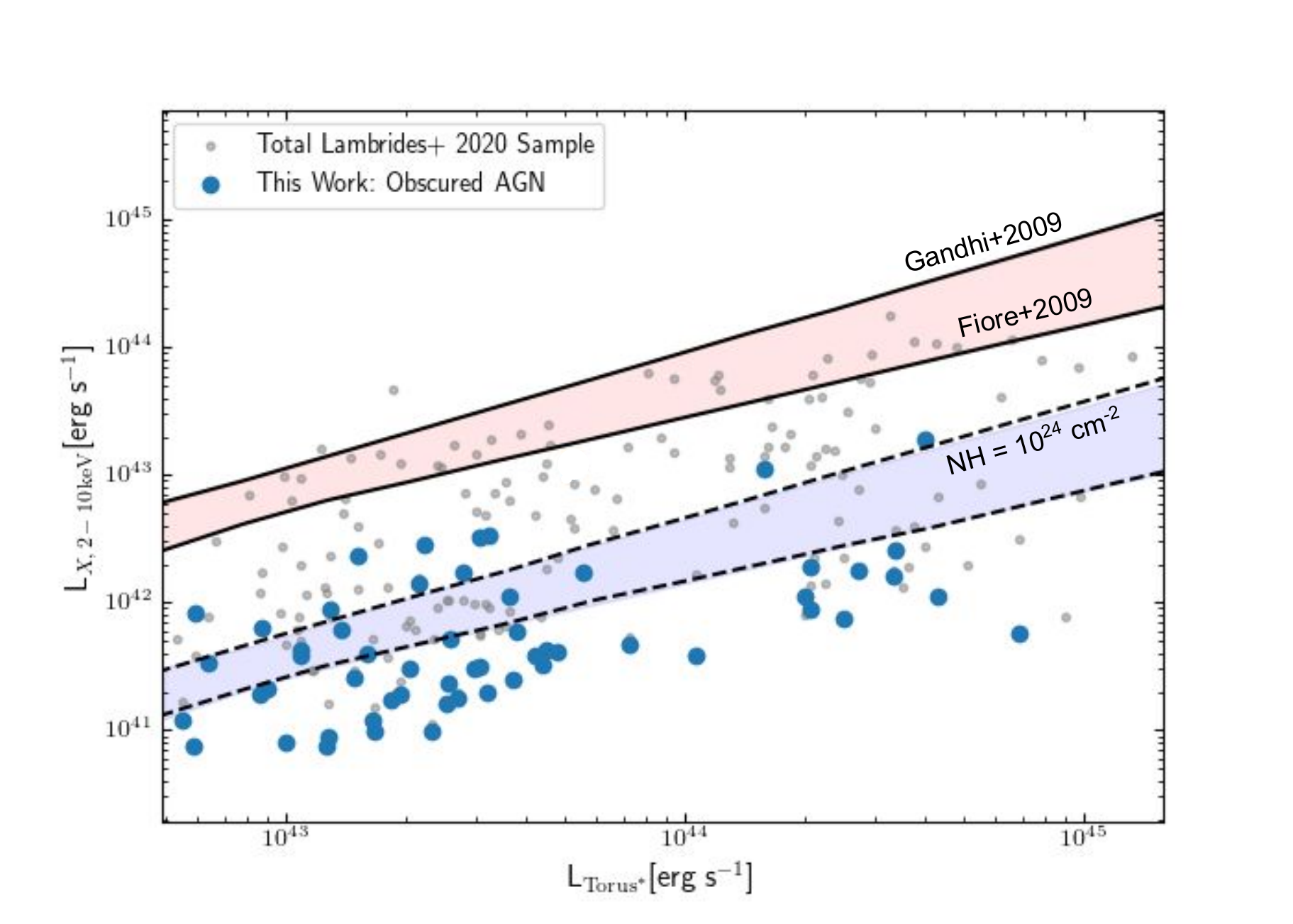}
	\caption{Non-absorption Corrected X-ray Luminosity vs Rest-Frame AGN MIR Luminosity: Obscured AGN candidates straddle or lie below the blue shaded region. As adapted by \citet{lansbury15}, the un-obscured region parameter space (red) indicates the range in intrinsic X-ray, 6 \micron\ AGN luminosity relationships between \citet{gandhi09} and \citet{fiore09}. The heavily obscured region (blue) indicates the same relationships but where the X-ray luminosity is absorbed by a column density of $N_\mathrm {H} > 10^{24}$ cm$^{-2}$ \citep{lansbury15}} 
\label{fig:nh_lx}
\end{figure*}

Directly observing X-ray bright obscured AGN with the \textit{Chandra X-ray Observatory} has been possible, especially at energies greater than $2$ keV where X-ray photons are less attenuated by the obscuring material. Generally, X-ray AGN are commonly selected in the literature as sources with intrinsic X-ray luminosities greater than the maximum luminosity one would expect from host-galaxy emission only (i.e $>10^{42}$ erg/s) and/or sources with enough X-ray photons in multiple energy bands to robustly model the X-ray spectrum. The latter condition is especially required to estimate the level of attenuation of the X-ray photons. In addition to X-rays, obscured AGN can also be identified in the mid-infrared (MIR) due to the dust reprocessing of the obscured UV light that emits from the central engine or through polarized scattered light \citep{houck05,stern12,mateos13}. The combination of wide and deep X-ray surveys with MIR multi-wavelength catalogues have greatly increased the samples of obscured AGN \citep[e.g.][]{stern05,donley12}. 

We derive our sample from the \citet{lambrides20}, hereinafter L20, lower luminosity obscured X-ray AGN catalogue. Utilizing the excellent wavelength coverage of the GOODS-South field, L20 analyzed the X-ray luminosities of AGN from the \textit{Chandra} 7Ms survey (7MsCDFS) in the context of the radio (VLA 1.4 GHz), optical grism spectroscopy (\textit{HST}-WFC3), high resolution optical/NIR imaging and photometry (\textit{HST}-ACS, \textit{HST}-WFC3IR), and NIR/MIR/FIR photometry (\textit{Spitzer} IRAC,\textit{Spitzer} IRS PUI, \textit{Spitzer} MIPS, \textit{Herschel} PACS). Using the absorption corrected 2-7 keV X-ray luminosities provided in the \citet{luo17} 7Ms catalogue, L20 derived an additional absorption correction factor to X-ray luminosities and thus to the N$_{H}$ of each object. This was done by measuring the offset of the \citet{luo17} luminosities from the X-ray luminosity required to be in agreement to within 2$\sigma$ of the \citep{stern15} empirical AGN X-ray to IR luminosity relationship where the IR estimate of AGN power is the rest frame IR luminosity between 3.6 \micron to 5.8 \micron. Using the IR excess in combination with X-ray and radio properties, L20 increased the number of identified obscured AGN in the 7MsCDFS catalog at 0.5 $<$ z $<$ 3 by 30\%, bringing the total number of 7MsCDFS obscured AGN with N$_{H} > 10^{23}$ cm$^{-2}$ to $\sim 100$. 

The 7Ms survey covers an area of $\sim$290 arcmin$^2$, and the L20 sample is distributed throughout this field. The Cosmic Assembly Near-IR Deep Extra-galactic Legacy Survey (CANDELS) \citep{candels} and 3D-HST \citep{skelton14} programs and resulting catalogues provide HST coverage for a portion of this field ($\sim$ 176 arcmin$^2$). To derive a suitable sample for this work, we first select the portion of the L20 sample that is within HST coverage using the mosaics provided by the 3D-HST\footnote{https://archive.stsci.edu/prepds/3d-hst/} \citep{koekemoer11,grogin11,skelton14}. 

Reliable X-ray-to-HST associations have been found for the CDFS catalogue in \citet{luo17} using the likelihood ratio technique presented in \citet{luo10} with the X-ray full-band derived coordinates. We use the X-ray counterpart F125W derived coordinates. The counterpart association described in \citet{luo10}, which takes into account positional uncertainties of the X-ray and F125W band and expected magnitude distribution of counterparts has a false-match probability $<$ 4\%. From the CANDELS+3DHST combined catalogue, the F125W band has a 5$\sigma$ limiting AB magnitude of 28.3. We test whether there is a statistical difference in the redshift, X-ray-luminosity and N$_H$ distributions of the AGN with HST coverage compared to the total L20 sample, and find that the null hypothesis cannot be rejected where the null hypothesis is that the distributions are identical (p$_{ADK}$ $>$ 0.25). The redshifts are provided in the \citet{luo17} \textit{Chandra} 7Ms X-ray catalogue: 46 are spectroscopic and 4 are photometric. In summary, we find a total of 50 obscured AGN out of the L20 obscured AGN sample with well covered ACS F435W, ACS F775W, and WFC3-IR F160W imaging data.



\subsubsection{X-ray and MIR Properties}

These 50 objects occupy a wide range of X-ray and MIR luminosities. The X-ray and MIR luminosities were derived in L20. The rest-frame MIR luminosity is used as an additional probe of AGN power and is defined between 3.2~\micron\ to 5.7~\micron. AGN torus emission dominates over MIR star-formation (SF) processes in this wavelength range which is especially pertinent for lower-luminosity, moderate-redshift AGN where other photometric MIR diagnostics may fail to capture these objects \citep{laurent2000,nenkova,kirkpatrick,lambrides}. In L20, the rest frame AGN MIR luminosity, referred to as L$_{Torus*}$, is calculating the photmetric luminosity of a single datapoint using the passband that most closely corresponds to the rest-frame wavelength range of interest. For the range of redshift spanned by our sample, the passbands used are the IRAC 8~\micron, IRS PUI 16~\micron\ and MIPS 24~\micron\ and for further detail on the MIR cross-matching and rest-frame luminosity calculation we refer the reader to the aforementioned paper. 

In \autoref{fig:nh_lx}, we show the non-absorption corrected X-ray luminosities  compared to the AGN luminosity in the MIR (L$_{Torus*}$). The red-shaded region corresponds to the un-obscured AGN region of the parameter space. This is defined by the range in intrinsic X-ray, rest AGN MIR luminosity relationships between two different X-ray to MIR relationships: \citet{gandhi09} and \citet{fiore09}. The \citet{gandhi09} relationship was derived from a local sample of type 1 AGN ($0.03 <$ z , $8\times 10^{41}$ erg/s $<$ L$_{X} < 4 \times 10^{43}$ erg/s), and decomposition of the nuclear 6 \micron\ luminosity was performed to minimize host-galaxy contamination. The \citet{fiore09} relationship was derived from a sample that spanned a larger redshift and X-ray luminosity range as compared to \citet{gandhi09} ($0.7 <$ z $< 2.2$, $3\times 10^{43}$ erg/s $<$ L$_{X} < 10^{45}$ erg/s), and did not include host-galaxy decomposition of the 6 \micron\ luminosity. Due to the inherent uncertainties of these relationships, instead of choosing a single empirical relationship, L20 chose a conservative approach and instead used both of these relationships to determine a region of the parameter space that corresponded to less obscured AGN. The heavily obscured region indicates the same empirical relationships but the X-ray luminosity is scaled down to represent a column density of $N_\mathrm {H} > 10^{24}$ cm$^{-2}$ \citep{lansbury15}. 

The blue points in \autoref{fig:nh_lx} comprise the heavily obscured AGN sub-sample from L20 with HST coverage and are not-significantly star-bursting as described in the previous section. In the next section we discuss the motivation and the removal of AGN host galaxies with starbursts.

\begin{figure*}
\centering
\includegraphics[scale=.9]{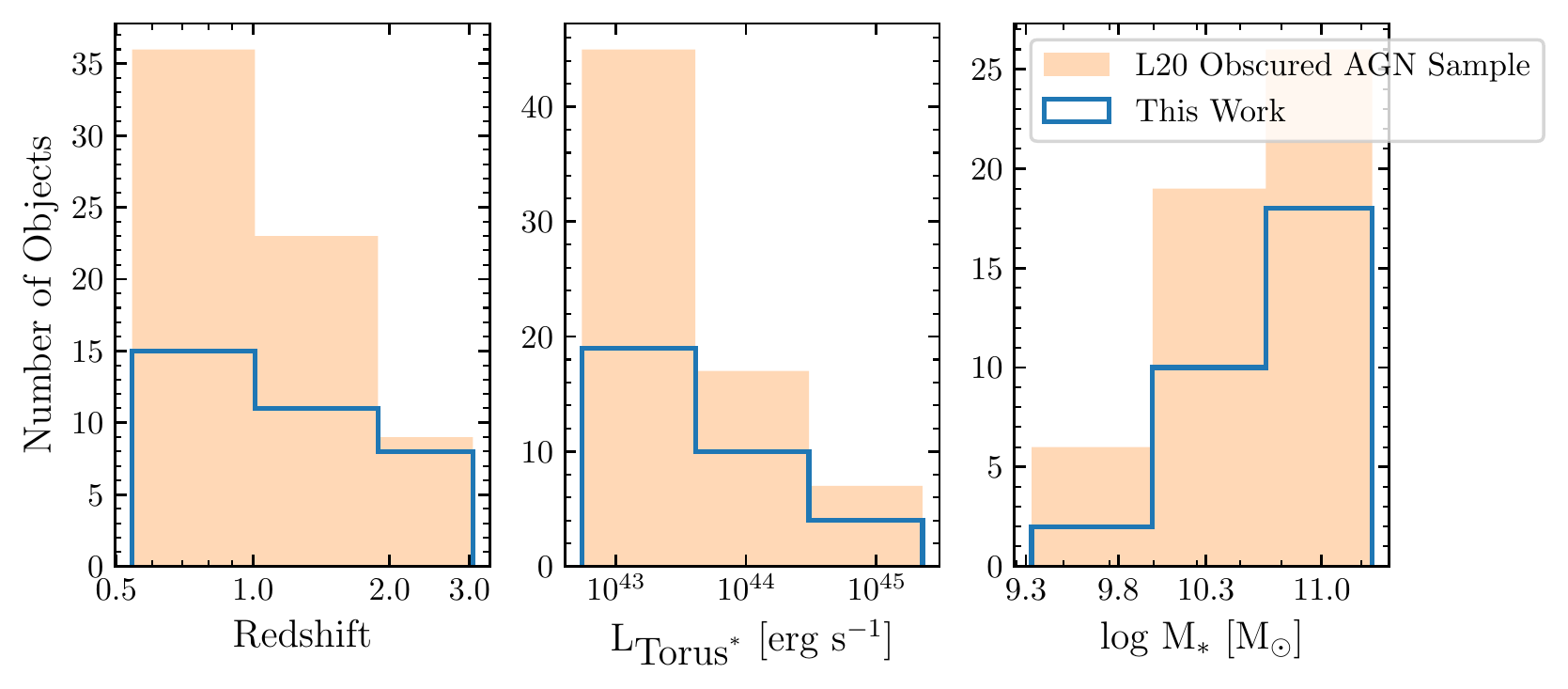}
	\caption{Parent Sample and This Work: Comparison of properties of the L20 Obscured AGN Sample to the non-starbursting, HST covered sub-sample used in this work.} 
\label{fig:sub_props}
\end{figure*}

\subsubsection{Removing Starbursts} \label{sec:rmsbs}
A multitude of theoretical and observational evidence has accumulated that potentially connect galaxy mergers and interactions to extreme bursts of star-formation or starbursts (SB) \citep{sanders96, hibbard96,hopkins06,davies2015,cortijo2017,pearson2019sbs, moreno2019}. The purpose of this work is to test the prediction that obscured AGN are more likely to be found in galaxies that are undergoing a significant merger. If there is a direct causal connection between mergers and star-formation and a star-formation rate (SFR) matched control sample is not used, an apparent secondary correlation between AGN and mergers can be induced. Thus, assessing the star-formation properties of the obscured AGN sample and the matched control sample is paramount. It is difficult to calculate robust star-formation rates of AGN host galaxies from photometry alone, and a careful analysis of the star-formation properties of the obscured AGN hosts is outside the scope of this paper due to the type of data in hand. Therefore, we identify sources that are likely undergoing the most extreme episodes of star-formation for a given stellar mass and redshift, and isolate them from the main sample. Due to the small number of obscured AGN with SB in their hosts, our main analysis will focus on the non-SB obscured AGN sample. The scope of this work is to test the hypothesis that the majority of obscured AGN are predominately triggered by significant galaxy mergers. In \autoref{sec:discussion}, we explore the merger properties of the SB-obscured AGN sample, and the implications of a SB-AGN-merger connection versus a non-SB-AGN merger connection. 

From this sample of 50 obscured AGN with HST coverage, we then select objects that are either likely to be on the star-formation main sequence, or quiescent. Utilizing the extensive wavelength coverage of the GOODS-S field, we calculate the position of the obscured AGN relative to the SF main-sequence for each galaxy's redshift and stellar mass. The stellar masses of the sample are given in the 3D-HST survey catalogue \citep{skelton14}. As described in \citet{skelton14}, these authors used the FAST code \citep{kriek09} to estimate the stellar properties of the entirety of the GOODS-S field. Due to the obscured nature of the AGN, the derived stellar masses are more robust than the other stellar properties estimated in the catalogue. 

As is stressed in \citet{skelton14}, the star-formation rates are uncertain when they are derived solely from optical- near-IR photometry. Since our obscured sample is heavily obscured (N$_H$ $>$ 5$\times 10^{23}$ cm$^{-2}$), the stellar masses are well constrained as they predominately depend on the rest-frame optical fluxes of the galaxies where there is negligible contamination from the central engine. The redshift range of our sources and the multiple HST band coverage allow for the rest-frame optical fluxes of our galaxies to be well measured. To estimate the SFR in our galaxies, we use the detections (or lack of) in the far-infrared (FIR). The FIR is a more un-biased indicator of star-formation than the MIR in AGN host-galaxies because the contribution from nuclear hot dust heated by the AGN contributes less than $<$ 20\% at $>100$ \micron\ even for the most powerful AGN  \citep{kirkpatrick2015,dai18,brown}. In this work, we estimate the SFR as traced by the 100 \micron\ and 160 \micron\ Herschel PACS band, utilizing the redshift information of the source and the SFR calibration provided in \citet{calzetti10}. The coverage and detection of the AGN sample at such wavelengths is discussed in L20. For objects with non-detections, we estimate the SFR using SFR$_{160 \micron} M_{\odot} yr^{-1}$ = L$_{160 \micron}/7 \times 10^{42}$  for $L_{160 \micron}$ $>$  2$\times$10$^{42}$ erg s$^{-1}$ $\sim 5.2 \times 10^{8}$ L$_{\odot}$. For the 31 non-detections, we estimate the SFR upper-limit by calculating the $L_{160 \micron}$ upper-limit using the 3$\sigma$ average depth limit of 2.7 mJy as presented in \citet{elbaz11}. 

We then use the SFR relation for main-sequence galaxies presented in \citet{schreiber2015} to calculate the SFR of the main-sequence galaxies at each object's mass and redshift. Starbursts are defined as 0.6 dex above the main-sequence population for a given stellar mass, SFR and redshift \citep{rodighiero2011}. Of the 50 obscured AGN, we remove from the sample the sources that are 0.6 dex or more above their main-sequence counterpart. This leaves the final non-SB obscured AGN sample with 40 objects.

\begin{figure*}
\centering
\includegraphics[scale=.9]{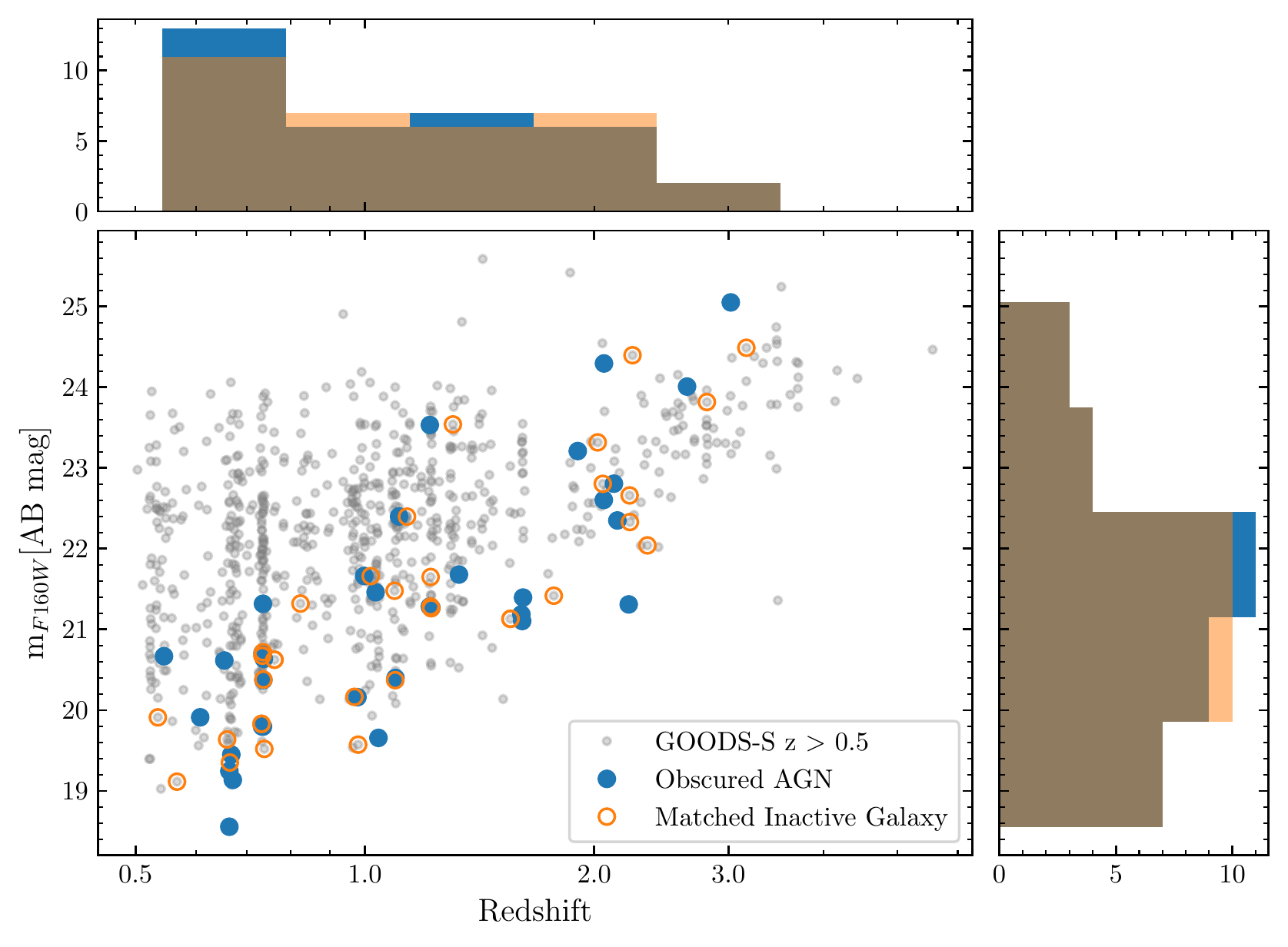}
	\caption{F160W AB magnitude versus Redshift: The grey points are the entirety of the 7Ms sample with HST coverage for z $>$ 0.5 in GOODS-South. The blue points are the redshifts and F160W magnitudes of the non-SB obscured AGN sample. The orange points are for the counterpart inactive-galaxy control sample.} 
\label{fig:obs_control_xmatch}
\end{figure*}

In \autoref{fig:sub_props}, we show a comparison of the redshift, L$_{\textrm{Torus}^{*}}$, and stellar mass distributions of the L20 parent sample compared with the limited sample used in this work. We calculate the k-sample Anderson-Darling mid-rank statistic between the redshift, L$_{\textrm{Torus}^{*}}$, and stellar mass distributions of the L20 sample of heavily obscured AGN to the sub-sample used in this work and find the null hypothesis cannot be rejected, and thus the sample used in this paper is representative of the obscured AGN found in the larger 7Ms survey.

\subsection{Control Sample}
Since the goal of this work is to measure any significant excess of mergers in the obscured AGN sample as compared to non-active galaxies, a control sample must be carefully selected to closely match the properties of the AGN hosts. We one-to-one match the non-SB obscured AGN to non-AGN galaxies (within $\Delta_{m_{F160W}} \pm 0.5, \Delta_{z} \pm 0.5 $) using the 3D-HST photometry catalogue \citep{skelton14} and spectroscopically secure redshifts from \citet{momcheva16}. If multiple galaxies satisfy the $\Delta_{m_{F160W}}$, $\Delta_{z}$ criteria, we select the galaxy with the smallest difference. The mean differences of $\Delta_{m_{F160W}}$ and  $\Delta_{z}$ between the non-SB obscured AGN sample and counterpart sample are -0.04 and -0.03, respectively. We choose only one counterpart galaxy per non-SB AGN to ensure the total sample is of reasonable size for visually classification. We use this non-AGN galaxy sample, herein called the control sample, to assess the presence of an obscured AGN-merger connection. 

Matching the control galaxy sample to the star-formation properties of the AGN host galaxies is a necessity. We remove starburst galaxies from the catalog as described in the previous section. In \autoref{fig:obs_control_xmatch}, we show the redshift and F160W magnitude distribution of the non-SB obscured AGN sample (blue points), control sample (orange points), and the entire z $>$ 0.5 GOODS-South field with starbursts and galaxies with photometric redshifts included (grey points). We find the distributions in redshift and magnitude are statistically indistinguishable between the non-SB obscured AGN and control sample: the null-hypothesis that the two samples are drawn from the same distribution in both z and Ks magnitudes cannot be rejected (p$_{ADK}$ $>$ 0.25). In \autoref{fig:obs_control_xmatch}, we include the total GOODS-S z $> 0.5$ sample to visually compare the region of the parameter space the non-SB obscured AGN and counterpart sample occupies to the in-active galaxies not matched to the non-SB obscured AGN sample.

To summarize, we cross-match a non-starbursting, HST covered sub-sample of the L20 obscured AGN catalogue to the CANDELS+3DHST combined HST GOOD-S catalogue \citep{skelton14,momcheva16}. The HST-covered L20 subsample consists of 40 non-SB obscured AGN with 34 spectroscopic redshifts and 6 photometric redshifts. The control sample consists of 40 non-active redshift, mag$_{F160W}$ matched counterpart galaxies all with spectroscopic redshifts.  The distributions of the non-SB obscured AGN sample and control sample distributed in redshift and F160W magnitude space are statistically identical with 0.5 $<$ $z$ $<$ 3.1 and 18.2 $<$ mag$_{F160W}$[AB] $<$ 25.1. 


\begin{figure*}
\centering
    \subfloat[\centering Obscured AGN: ID 642 from \citet{luo17}]{{\includegraphics[width=5cm]{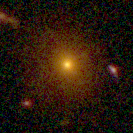} }}%
    \qquad
    \subfloat[\centering Counterpart Non-AGN Galaxy: ID 18169 via \citet{skelton14}]{{\includegraphics[width=5cm]{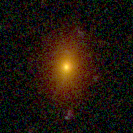} }}%
    \caption{Example of Sample Imaging: Obscured AGN RGB Image at z = 0.7 (a) with its z, F160W matched non-AGN galaxy counterpart (b)}%
\label{fig:obs_con_eg}
\end{figure*}

\subsection{HST Datasets} \label{sec:imaging}

The number density of obscured AGN is inferred to peak between 1 $<$ $z$ $<$ 2 \citep[e.g.][]{gilli07,aird2015}. Beyond optical z$\sim$ 1 imaging, surveys begin to probe the rest-frame UV morphologies of galaxies. This is useful for probing the most active regions of unobscured SF, but may miss the gaseous and stellar features associated with merging systems (i.e shells, disk asymmetry). An additional complication with morphologically analyzing z $>$ 1 galaxies is the increasing incidence of foreground and background galaxies near the region of the object of interest. Color images are helpful in determining whether a close pair is a random superposition of galaxies, or two galaxies at the same redshift. Thus, we need multiple optical/UV imaging bands at similar depth in order to assess the merger status of a $z$ $>$ 1 galaxy. 

In this study, we use the 3D-HST reduced and combined GOOD-S mosaics \citep{skelton14}. We make a 6" x 6" cutout centered on the X-ray coordinates for the obscured AGN sample, and the 3D-HST coordinates for the control sample. Each of the postage stamps were individually inspected to ensure no prominent image artifacts were in the cutouts. All the postage stamps of the obscured and control sample in this study are made publicly available.\footnote{erinilambrides.com/morphology\_of\_obscured\_agn}.

\subsection{Mock Galaxy Sample}
An important aspect of merger classification studies is the major uncertainty associated with the accuracy of human classifiers. By accuracy, we mean the ability of each person to correctly identify mergers and to disentangle them from random super-positions, asymmetries in galaxy structure not due to tidal interactions, and relaxed morphologies. Even if the classifiers are experts and proper statistical analysis is performed to remove outliers \citep[e.g. using trimmed means as in][]{chiaberge15}, a bias can still be present. L21 found that  when  using  one  of  the  most  standard  statistical  implementations used to calculate the merger fraction in the literature, the effective bias due to humans is dependent on the intrinsic merger fraction of a given sample. The implications of this result cast doubt in the sole usage of a control sample as justifiable means to encapsulate human bias. L21 proposed a method of quantifying and accounting for merger biases of individual human classifiers and incorporated these biases into a full probabilistic model to determine the merger fraction of a population, and the probability of each individual galaxy being in a merger. In \autoref{sec:methods_summary}, we summarize the formalism and results of L21 on the definition and effect of the bias introduced by human classifiers in addition to the statistical framework used to infer the merger fraction of a sample.  

In L21 we introduced a new method to calibrate the accuracy of human classifiers. An estimate of the classifier’s accuracy was used as a prior in determining the merger fraction of a galaxy sample. The accuracy priors were determined using simulated images from the VELA cosmological simulations \citep{ceverino14, snyder15,simons19}. As shown in L21, 50 mock images in three different bands (two in the optical and one near-IR) were produced with the appropriate amount of Poisson noise to simulate the real data-sets used in this research. The mock galaxy sample has the same redshift distribution as the non-SB obscured AGN sample used in this work. With their origin hidden, these simulated observations were also classified by each of the co-authors in this work, and for more details on the construction of the mock images we refer the reader to L21.


\section{Determining a Data Driven Merger Fraction} 

At z$>$1, it becomes more difficult to accurately assess the merger state of a galaxy as faint merger signatures may be undetectable \citep{lotz}. Despite the great potential of automated methods such as deep-learning for merger identification, there currently is no tool that is robust enough to handle the diverse presentations of merging galaxies in the earlier Universe \citep{pearson2019sbs}. Visual human classification is the most commonly employed method used to identify moderate samples of merging galaxies at $z > 1.0$, but rarely if ever do the authors of these studies attempt to control for human bias in morphological studies aside from the usage of a control sample. 

In L21, we used simulated and observed data-sets, to create and validate a data-driven merger fraction probability model, where the merger fraction is defined as the fraction of galaxies within a given sample undergoing a significant merger. For the observed data-sets, we used real human classifications on a sample of mock images with known truth values derived from cosmological simulations. We found that the bias introduced from human classification is dependent on the intrinsic merger fraction of the population, and not accounting for this bias can drive the resulting merger classification rates to be significantly different from the intrinsic truth. The statistical framework posed in L21 accounts for the merger classification biases of individual human classifiers, and these biases are then incorporated into a full probabilistic model to determine the merger fraction of a population and the probability of an individual galaxy being in a merger. In this section, we describe how the human classifications of the non-SB obscured AGN and in-active galaxy counterpart sample were collected and analyzed using the L21 framework \footnote{The full source code of the likelihood maximization can be found here: \url{https://github.com/elambrid/merger_or_not}}. 

\subsection{Object Classification Method}

We developed a website where classifiers could assess the morphologies of the non-SB obscured AGN sample, the control sample, and the mock galaxy sample without knowing which sample an object came from. The Morphology of Obscured AGN (or MOOAGN) classifying framework comprises the entirety of the sample: 40 obscured AGN, 40 matched inactive galaxies, and 50 mock galaxies. We also provided a demo survey of 5 objects (not used in the MOOAGN sample) to give the classifiers a reference framework of the classification options and data quality. At the end of the demo survey we give some example justifications of why one would classify an object as such, (see \autoref{sec:appendix} for demo survey and example classifications). Ultimately, for our analysis we use only two morphological classes: merging and not merging. Due to the difficulty in constraining merger stage and mass ratio given the data in hand, further morphological sub-divisions would yield potentially less accurate results. Nonetheless, when the human classifiers are presented with the images they are given multiple morphological divisions to choose from. This is to not only aid in the human classification process, but also to take the most conservative approach of testing for a merger excess in non-SB obscured AGN host galaxies. We assume that any system with obvious merging features observed at these redshifts must be significantly merging systems. If we are incorrect with this assumption, then the merger fraction would be lower for major-merging systems. After the sample was classified, the divisions were folded back into the two morphological classes of merging or not merging. The five classification options given to the human classifiers are as follows: 

\begin{enumerate}
    \item Merging: Major (approximately similar size) - On-going interaction. This is prior to coalescence i.e two distinct interacting galaxies of similar size. Features for this classification can include tidal tails with distinct galaxy pairs, enhanced star formation and morphological distortion along the closest axis of approach between two pairs.
    \item Merging: Minor (approximately $<$ 1:4 size ratio) Similar to the major merger classification with exception of size. If a galaxy pair has evidence of interaction, and one of the bodies is roughly less than a 1/4 the size of the larger galaxy it is classified as a minor interaction.
    \item Disturbance: Major: This is intended to capture galaxies that have coalesced within 100 Myrs. Features can include highly irregular gas/stellar morphologies and tidal tails with only one distinct central bulge
    \item Disturbance: Minor - This is intended to capture galaxies that are slightly irregular, yet are indistinguishable from internal processes that could cause the irregularity i.e star-forming clumps, disk instabilities. 
    \item No Evidence of Merger/Interaction. 
\end{enumerate}

Examples of galaxies fitting the above criteria are shown in \autoref{sec:appendix}. We then collate the classifications of our fourteen human classifiers of all 130 objects on the MOOAGN sample.

\subsection{Calculating the Merger Fraction Likelihood} \label{sec:methods_summary}

As previously mentioned, even among experts, it is difficult to accurately characterize whether a galaxy is undergoing a merger or is isolated. Because of this, it is inevitable that any given classifier will obtain a merger fraction that is different from another classifier's assessment. For example, one may be more inclined to classify objects as mergers even if the objects display minor disturbances unrelated to galaxy encounters. L21 assumes that the bias of human classifiers can be quantified in terms of their accuracy in correctly classifying an intrinsically merging galaxy as a merger (and an intrinsically isolated system as isolated). Previous works have assumed that the effect of this bias on independent galaxy samples is similar (i.e if the same set of humans are classifying a science and a control sample the assumption is that the bias due to human classification is equally present in both samples). Thus due to this assumption, and to account for other un-quantified biases such as those potentially introduced during the method of selecting the science sample in the first place, most merger studies do not report absolute merger fractions of a specific population but rather compare the merger fraction of the science sample to a well-justified control sample. The control sample in this context is any sample of sources that lacks the key feature that defines the science population in question, but shares any relevant properties that might be correlated with the morphology or the presentation of the morphology of an object (i.e redshift, stellar-mass, SFR etc).  Though, as shown in L21, if the underlying merger fraction of the two populations (i.e science and control) are significantly different, this human bias will not be evenly applied. Therefore, the bias introduced by using human classifiers will still be present in any statistical comparison between the merger fractions of the science and control sample.   


Summarizing the L21 characterization of this bias, if one is shown a merging (or isolated) galaxy, they will classify the galaxy correctly with probability $r_M$ (or $r_I$). Therefore, if somebody is shown $N_M$ intrinsic mergers and $N_I$ intrinsic isolated galaxies, on average they will measure
$\hat N_M=r_M N_M + (1-r_I)N_I$ mergers. The inclusion of the $(1-r_I)N_I$ term represents the amount of galaxies that were incorrectly classified as isolated and are truly mergers.

Using the formalism of $r_M$ (or $r_I$) to characterize the bias of human classifiers, L21 shows that the use of relative significance between comparing the merger fractions of the science and the control sample does not remove this issue. By re-writing the measured $\hat N_{M}$ and measured $ \hat N_{M,c}$ in terms of the measured merger fraction for each sample and the intrinsic value of $N_{M}$ and $N_{M,c}$  in terms of the intrinsic merger fraction $f_{M}$ of each sample and taking the difference:
\begin{align}
    \langle \hat f_M\rangle
    &=r_Mf_M+(1-r_I)(1-f_M)
    \label{eq:fm_hat}
    \\
    \langle \hat f_{M,c}\rangle &=
    r_Mf_{M,c}+(1-r_I)(1-f_{M,c})
    \\
    \langle \Delta\hat f_M\rangle &=
    \langle \hat f_M\rangle - \langle \hat f_{M,c}\rangle
    \\
    \langle \Delta\hat f_M\rangle  &=
    r_M\Delta f_M
    -(1-r_I)\Delta f_M
    \nonumber
    \\
    &=\Delta f_M[r_M+r_I-1]
\end{align}

they find the difference between the measured merger fractions of the two samples is still dependent on the intrinsic merger fraction of each sample.

Using the merger fraction likelihood algorithm presented in L21, we are able to infer the underlying merger fraction by using a novel technique to quantify the bias of each individual classifier. We then optimally combine the individual classifier uncertainties with the individual classifications of each galaxy in the sample. In the merger fraction statistical model presented in L21, classifier accuracy is a nuisance parameter that can be marginalized over. Further details on the construction of the algorithm can be found in the aforementioned work. We briefly summarize the algorithm here. 

A respondent $i$ is shown a true merger, and they classify it as a merger with probability $r_M$, or classify it as an isolated galaxy with probability $1-r_M$. Conversely, if the respondent is shown a true isolated galaxy, they will say it is a merger with probability $1-r_I$ or say it is isolated with probability $r_I$. Thus respondent $i$ classifies $j$th galaxy $G_{j}$ with classification $m$ as
\begin{equation}
    p(m_i\mid G_j)
    =
\begin{cases}
    r_M &   m_i=G_j=\mathrm{merger}\\
    1-r_M&m_i\neq G_j=\mathrm{merger}\\
    r_I&m_i=G_j=\mathrm{isolated}
    \\
    1-r_I&m_i\neq G_j=\mathrm{isolated}
    \end{cases}
\end{equation}

 \begin{figure*}[hbt]
\centering
\includegraphics[scale=.9]{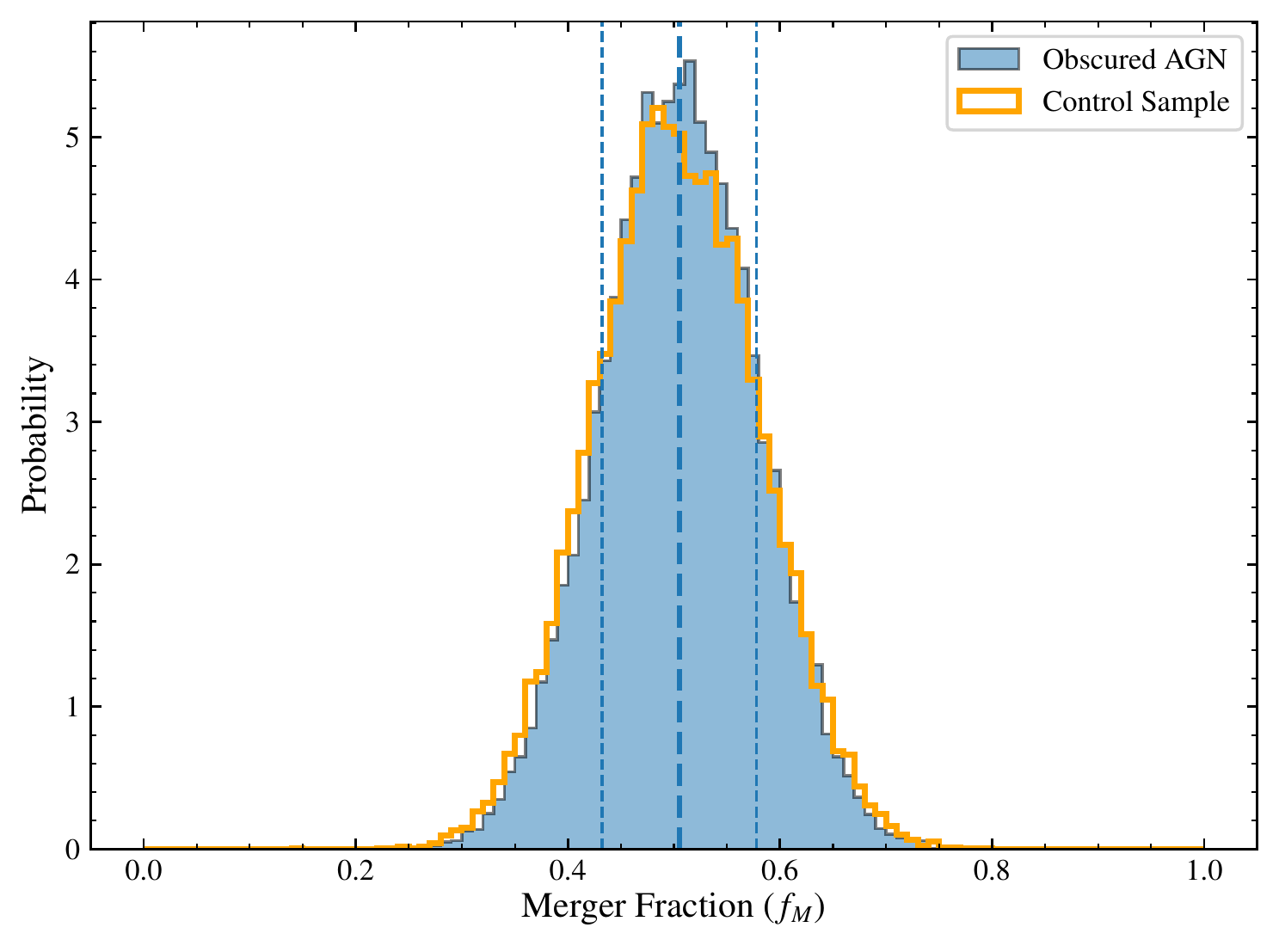}
	\caption{Merger Probabilities of non-SB Obscured AGN and Inactive Galaxy Control Sample: We use the method presented in section 3.2 to calculate the probabilities of each individual galaxy in the non-SB obscured AGN and control samples being in a merging system. The blue distribution is merger fraction distribution of the non-SB obscured AGN sample, and the orange outlined, un-filled distribution is the merger fraction of the control sample. The blue line is centered at the mean of the obscured AGN distribution, with the blue dashed lines representing the 85th percentile. There is no significant difference in the merger fractions.} 
\label{fig:merger_iso_prob}
\end{figure*}

The likelihood of the classifications of a single galaxy by multiple classifiers given a merger fraction and classifier accuracies can be written as
\begin{equation}
\begin{split}
    p(\{m_i \} \mid \{r_i\}, f_M) = f_M \prod_{i} p(m_i\mid G_{j}={\mathrm M}) \\ 
    + (1-f_M) \prod_{i} p(m_i\mid G_{j}={\mathrm I}).
\end{split}
\end{equation}
where $f_M$ is the merger fraction of a given population and index $i$ corresponds to an individual classifier. In this expression, the true nature of the galaxy in question is marginalized out. Expanding to multiple galaxies, we get the likelihood for the classifications of a collection of galaxies:

\begin{equation}
    p(\{m_{ij}\} \mid \{r_i\}, f_M) = \prod_{j} p(\{m_{ij}\} \mid \{r_i\}, f_M).
\end{equation}
Multiplying this likelihood by a prior on the merger fraction and, if the classifier accuracies are not held fixed, by a prior on accuracies gives the unnormalized posterior probability distribution function for this model. If we wish to recover the probability that a particular galaxy is a merger, we can use the expression:
\begin{equation}
    p(G = M \mid \{m_i\}, \{r_i\}, f_M) = \frac{f_M \prod_i p(m_i \mid G=M)}
    {p(\{m_i\} \mid \{r_i\}, f_M)}.
\end{equation}
The probability that this galaxy is isolated is the complement of this expression. This expression is evaluated with an informative prior on the accuracies $r_M$, $r_I$. The prior is determined from the classifications of the mock galaxies, which have known merger states. We refer the reader to L21 for further details on the derived $r_M$, $r_I$ classifications. The same set of human classifiers were used in both L21 and this work, and the mean prior of $r_M$  $r_I$ is 0.74 and 0.63 respectively. 

The strength of this method is its internal consistency: given a set of observed mergers, the likelihood is maximized when a value of $f_M$ shown to all classifiers is most plausible, given a prior on classifier accuracies and the individual classifications of each galaxy. For example, in studies that determine the merger fraction of a population from a set of galaxies classified by a set of human classifiers, the merger fractions from each classifier are collated and the error treatment uses the standard binomial statistics. In this scenario, it is possible for classifiers to identify a similar number of mergers, but be in disagreement with each other on the classification of individual objects. This lack of inter-classifier agreement would not be encapsulated in the standard error treatment. In the method utilized in this work, the determination of the most plausible $f_M$ requires determining the most plausible classification of each individual galaxy.   

The likelihood function of a given galaxy having a specific morphological classification requires a robust statistical description of a human classifiers accuracy in assessing both merging and isolated systems. In the previous step, where we maximize the likelihood of a population's merger fraction, our algorithm also maximizes the likelihood of an individual galaxy's classification. This allows for deeper data exploration on galaxy samples that are normally too small to do anything but population averages. 

\begin{figure*}
\centering
\includegraphics[]{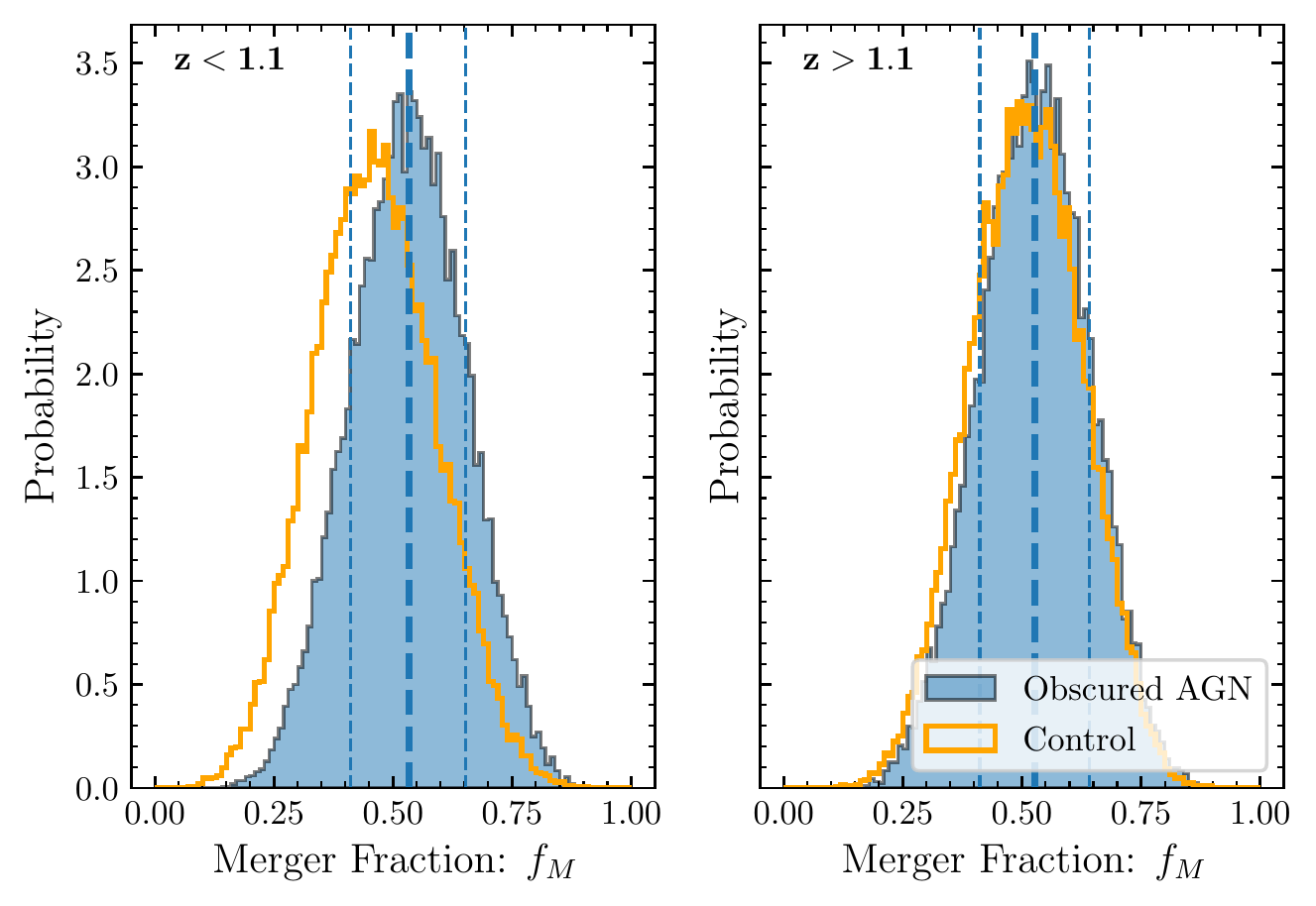}
	\caption{Merger Probabilities of non-SB Obscured AGN and Inactive Galaxy Control Sample in Two Redshift Bins: The dark blue filled histogram is the merger probability distribution of the non-SB obscured AGN sample and the un-filled orange histogram is the matched in-active galaxy sample. The left most plot represents objects in the lower 50\% of the redshift distribution (0.5 $<$ z $<$1.1 (20 objects), and the right-most plot the merger probability distributions for the objects in the upper 50\% of the sample redshift distribution (1.1 $<$ z $<$3.5).} 
\label{fig:redshift_2bins}
\end{figure*}

\section{The non-SB Obscured AGN Merger Fraction}

We first present the merger fractions of the non-SB obscured AGN and control sample without taking into account the accuracies of the human classifiers. We take the mean number of galaxies classified as either a major merger, minor merger, or majorly disturbed system from each of the fourteen classifiers. We also report the binomial confidence interval at the 68\% level, or $1\sigma$, using the Jeffreys interval, a Bayesian application to the binomial distribution \citep{jeffrey}. The merger fraction and corresponding 1$\sigma$ error of the non-SB obscured AGN sample is 0.59 $^{+0.06}_{-0.10}$. For the control sample, the merger fraction and 1 $\sigma$ error is 0.53 $^{+0.06}_{-0.11}$. The merger excess of non-SB obscured AGN over a matched inactive control sample is 1.1 $^{+0.3}_{-0.2}$. Thus, using the standard binomial method, the control sample and obscured sample are not statistically separable. Yet, as shown in L21, the only instance in which the relative comparison of two merger fractions using the standard binomial method is not biased due to human classification is when the two samples being compared have the same intrinsic merger fraction. Since we do not know \textit{a priori} the intrinsic merger fractions of the non-SB obscured AGN and control sample, we must use our newly derived method to estimate the merger fraction.

\begin{figure*}
\centering
\includegraphics[scale=.9]{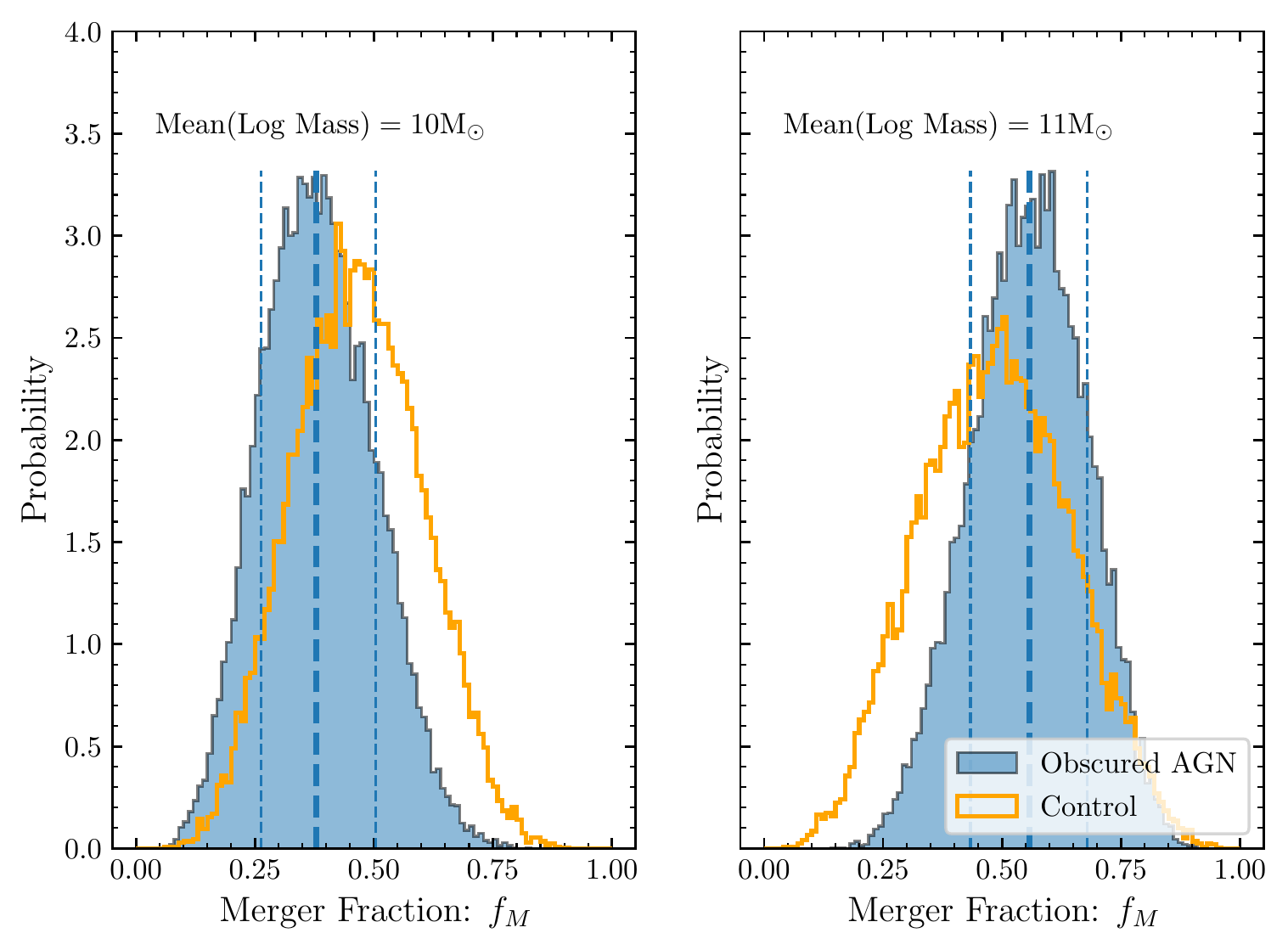}
	\caption{Merger Probabilities of non-SB Obscured AGN and Matched Inactive Galaxy Control Sample in Two Stellar Mass Bins: The dark blue filled histogram is the merger probability distribution of the obscured AGN sample and the un-filled orange histogram is the matched in-active galaxy sample. We split the non-SB obscured AGN sample on the median log stellar mass: 9.32 $<$ log (M$_{*}$ [M$_{\odot}$]) $<$10.7 (20 objects), 10.7 $<$ log (M$_{*}$[M$_{\odot}$]) $<$11.32 (20 objects). The left most plot represents the merger probability distributions of the lower stellar mass bin (log(M$_{*,\textrm{mean}}$) = 10 M$_{\odot}$), and the right-most plot the merger probability distributions for the higher stellar mass bin (log(M$_{*, \textrm{mean}}$) = 11 M$_{\odot}$).} 
\label{fig:stellarmass_2bins}
\end{figure*}


Thus, we use the merger fraction likelihood framework presented in L21 and summarized in \autoref{sec:methods_summary} to simultaneously calculate the probability of the merger fraction of each sub-sample, the probability distribution of each classifiers accuracy in measuring merging and isolating systems, and the probability of each individual galaxy being in a merger. In \autoref{fig:merger_iso_prob}, we report the merger fraction probability distribution of the non-SB obscured AGN and control sample being in an merging system. The y-axis probabilities are normalized such that the area under the distribution curve is equal to one. We find the non-SB obscured AGN sample has a merger fraction probability of 54\%$\pm 8\%$, and the inactive galaxy control sample is found to have a 53\%$\pm 9\%$ mean probability of being in a merger. 


The main result of our work is as follows: The obscured AGN merger fraction is statistically indistinguishable from the control sample merger fraction ( $<$1 $\sigma$). 

In the following sub-sections, we explore whether an intrinsic difference exists between the merger state of obscured and inactive galaxies as a function of various galaxy properties. We test the extent of the dependence on merger probability on different galaxy and AGN properties by simply splitting the sub-sample along the 50th percentile (or on either side of the median) of the property being explored. We do this to have enough objects in each bin to keep the error on the sub-sample size small enough for meaningful comparison and to minimize assumptions on bin width.

\begin{figure}
\centering
\includegraphics[]{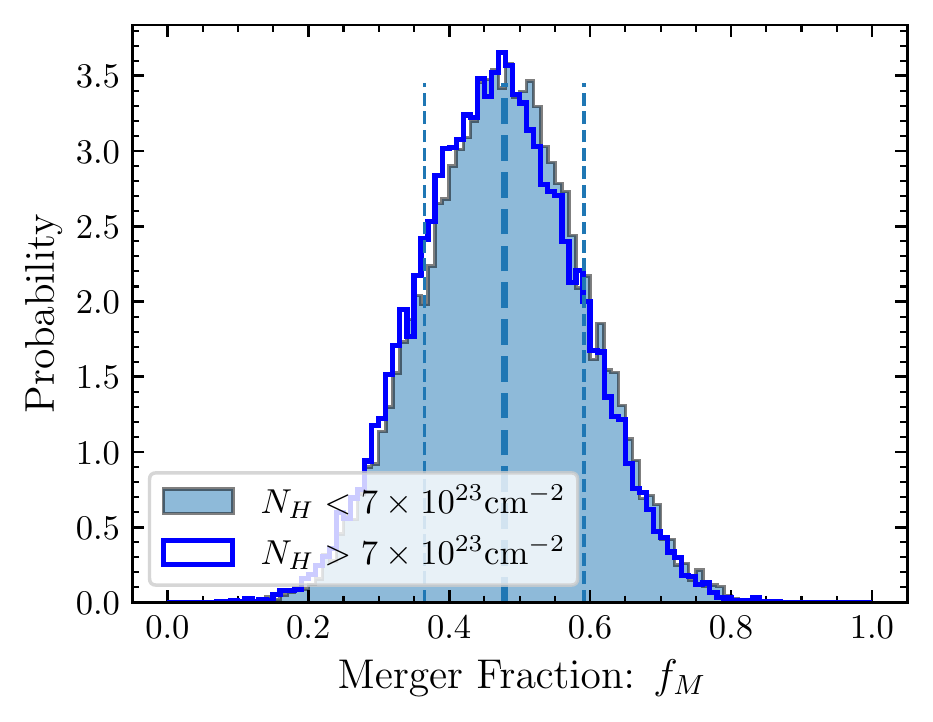}
	\caption{Merger Probabilities of non-SB Obscured AGN in two N$_H$ bins: The light blue filled histogram is for non-SB obscured AGN sample with objects with N$_{H} < $ 7$\times 10^{23}$ cm$^{-2}$, and the dark blue un-filled histogram for non-SB obscured AGN with N$_{H} > $ 7$\times 10^{23}$ cm$^{-2}$} 
\label{fig:NH_2bins}
\end{figure}

\subsection{Redshift Dependence}

We first compare whether there is a difference in the merger fractions as a function of redshift. We split the non-SB obscured AGN sample along the median, 0.5 $<$ z $<$1.1 (20 objects), 1.1 $<$ z $<$3.5 (20 objects), and split the control sample along those same bin definitions (the redshift median of the control sample is also 1.1, 20 objects in each bin respectively). In \autoref{fig:redshift_2bins}, we show the merger probabilities of the non-SB obscured AGN and control sample for each redshift bin. For the lower redshift bin, we find f$_M$ = 0.42 $\pm$ 0.11 and f$_M$ = 0.44 $\pm$ 0.12 for the non-SB obscured AGN sample and matched control sample respectively. For the higher redshift bin, we find f$_M$ = 0.51 $\pm$ 0.11 and f$_M$ = 0.49 $\pm$ 0.12 for the non-SB obscured AGN sample and matched control sample respectively. We do not find any statistical difference between the non-SB obscured AGN sample and the control sample for either the lower redshift or higher redshift bin ($< 1 \sigma$ difference).

\subsection{Galaxy Stellar Mass Dependence}

We next explore if there is any difference in the merger probabilities between non-SB obscured AGN and the control sample that is dependent on stellar mass. In \autoref{fig:stellarmass_2bins}, we again split the non-SB obscured AGN sample on the median log stellar mass: 9.32 $<$ log (M$_{*}$ [M$_{\odot}$]) $<$10.7 (20 objects), 10.7 $<$ log (M$_{*}$[M$_{\odot}$]) $<$11.32 (20 objects). Using the same bin widths, we split the control sample (20 objects). For the lower mass bin, we find f$_M$ = 0.39 $\pm$ 0.19 and f$_M$ = 0.45 $\pm$ 0.22 for the non-SB obscured AGN sample and matched control sample respectively. For the higher mass bin, we find f$_M$ = 0.55 $\pm$ 0.15 and f$_M$ = 0.49 $\pm$ 0.28 for the non-SB obscured AGN sample and matched control sample respectively. We find that again the difference between the non-SB obscured AGN sample and the control sample is not statistically significant ($< 2 \sigma$ difference).  



\begin{figure}
\centering
\includegraphics[]{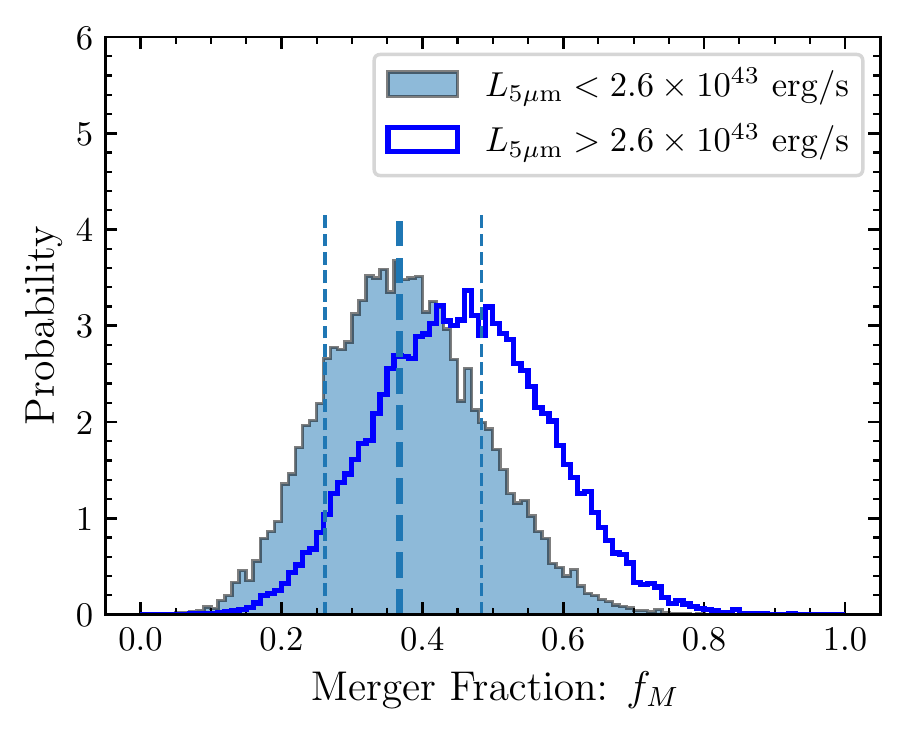}
	\caption{Merger Probabilities of non-SB Obscured AGN in two L$_{\textrm{Torus}^{*}}$ bins: The light blue filled histogram is for non-SB obscured AGN sample with objects with L$_{\textrm{Torus}^{*}} < $ 2.6$\times 10^{43}$ erg s$^{-1}$, and the dark blue un-filled histogram for non-SB obscured AGN with L$_{\textrm{Torus}^{*}} > $ 2.6$\times 10^{43}$ erg s$^{-1}$} 
\label{fig:L5mu_2bins}
\end{figure} 

\subsection{Dependence on Obscuration and AGN Power}

We then test whether there is any differences in merger probabilities for different levels of AGN obscuration and/or AGN power. In \autoref{fig:NH_2bins}, we split the non-SB obscured AGN sample along the median of obscuration to produce two bins of less obscured AGN (22 objects) versus more obscured AGN (25 objects). For the lower and higher N$_{H}$ bin, we find f$_M$ = 0.48 $\pm$ 0.11 and f$_M$ = 0.47 $\pm$ 0.11 respectively. We do not find a significant difference amongst the extremely obscured objects versus the moderately obscured objects, as the merger fractions are consistent with each other better than 1$\sigma$. As an additional test, we compare the lower and higher N$_{H}$ bins against each of their respective matched control samples. For the lower N$_{H}$ bin, we find f$_M$ = 0.48 $\pm$ 0.11 and f$_M$ = 0.46 $\pm$ 0.10 for the non-SB obscured AGN sample and matched control sample respectively. For the higher N$_{H}$ bin, we find f$_M$ = 0.47 $\pm$ 0.13 and f$_M$ = 0.43 $\pm$ 0.25 for the non-SB obscured AGN sample and matched control sample respectively.

We also test if there is a difference amongst the more powerful AGN in our sample versus less powerful AGN. As in L20, we use L$_{\textrm{Torus}^{*}}$ a rest-frame 5$\micron$ luminosity indicator, to probe AGN power. In \autoref{fig:L5mu_2bins}, we show the merger probabilities of the non-SB obscured AGN split along the median value of L$_{5\micron}$: 5.5$\times 10^{42}$ $<$ L$_{5\micron}$ (ergs/s) $<$2.7$\times 10^{43}$ (32 objects), 2.7$\times 10^{43}$ $<$ L$_{5\micron}$ (ergs/s) $<$ 2.3$\times 10^{45}$ (34 objects). For the lower and higher L$_{5\micron}$ bin, we find f$_M$ = 0.37 $\pm$ 0.25 and f$_M$ = 0.47 $\pm$ 0.31 respectively. For the 5$\micron$ rest-frame luminosity values we probe in our sample, we do not find a significant difference between the two bins of non-SB obscured AGN, as the merger fractions are consistent with each other better than 2$\sigma$. We then compare the lower and higher L$_{5\micron}$ bins against each of their respective matched control samples. For the lower L$_{5\micron}$ bin, we find f$_M$ = 0.49 $\pm$ 0.12 and f$_M$ = 0.47 $\pm$ 0.10 for the non-SB obscured AGN sample and matched control sample respectively. For the higher L$_{5\micron}$ bin, we find f$_M$ = 0.51 $\pm$ 0.12 and f$_M$ = 0.48 $\pm$ 0.16 for the non-SB obscured AGN sample and matched control sample respectively.

\section{Discussion} \label{sec:discussion}

In terms of merger fraction, we do not find any significant difference between our non-SB obscured AGN sample and a redshift, F160W, non-starbursting non-AGN galaxy sample. This is in tension with both theoretical and observational works that place heavily obscured AGN within a major-merger-driven evolutionary paradigm. 
It has been speculated that the AGN-merger connection may have been systematically missed due to poor sampling of obscured AGN \citep{kocev15}. \citet{kocev15} were amongst the first to attempt a careful investigation of such a relationship, by selecting one of the largest samples of obscured AGN of its time using multiple deep-field X-ray data-sets. However, differently from our work, they only used one HST NIR band (F160W), employed a smaller number of human classifiers (2), and their statistical analysis did not consider the biases we work to address here. Additionally, the control sample in \citet{kocev15} consisted of un-obscured X-ray selected AGN. They were selected to match their obscured sample in both redshift and X-ray luminosity only. Conversely, in this work our control sample consists of inactive galaxies. This is important because un-obscured AGN may have a significant un-resolved point-like component in their images, thus making morphological classification of and estimation of the stellar properties of host galaxies with bright point-sources extremely difficult. Interestingly, as noted by these authors, when they remove the sources with point source morphologies, the significance of the merger excess in the heavily obscured AGN sample drops from $3.8\sigma$ to $2.5 \sigma$. 

Another significant difference is the star-formation properties were not determined prior to control sample creation. As mentioned in Section \ref{sec:obs_sample}, both theoretically and observationally there is a strong association between mergers and starbursting galaxies \citep{sanders96,veilleux09,kartaltepe2012,rodriguez2019}. Many AGN studies focusing on the morphology of AGN host galaxies do not properly remove starburst galaxies from their samples. It is in fact very difficult to adequately take this into account. Since large samples of AGN with deep optical/UV imaging at the redshift distribution probed in this work usually lack the required high S/N IR spectra to accurately de-tangle the SF and AGN contribution to the IR. If the AGN sample contains more (or less) star-bursting galaxies than the control sample, a non-causal merger excess (or deficit) can be found in the obscured sample. 

\begin{figure}
\centering
\includegraphics[]{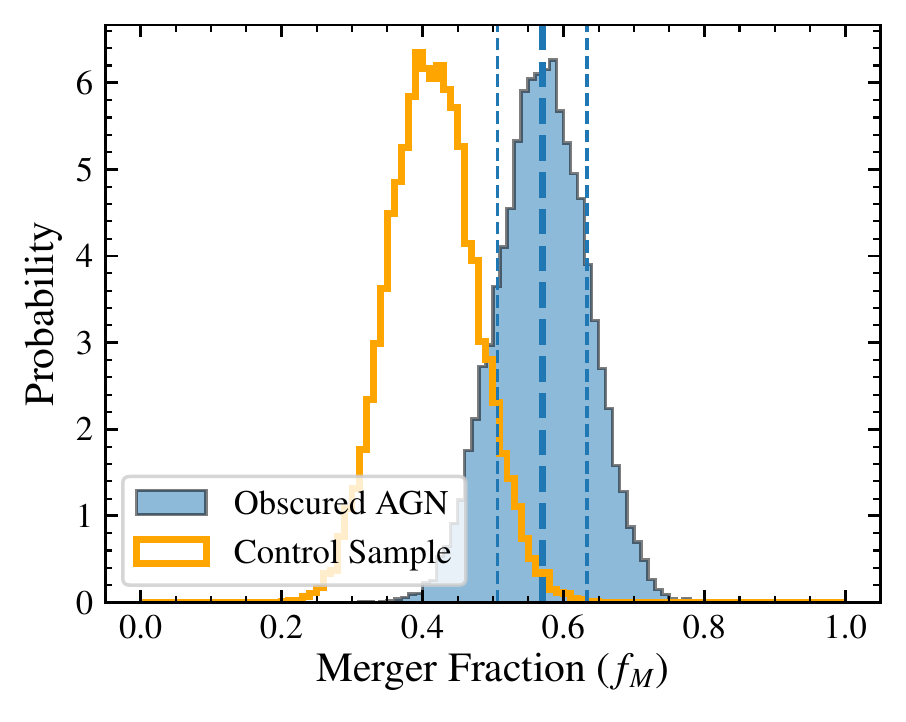}
	\caption{Including obscured AGN with hosts undergoing a star-bursts reveals the importance of including SF properties in the counterpart sample creation.} 
\label{fig:wstarbursts}
\end{figure}

To check if the uneven inclusion of starburst galaxies would generate any significant bias on our results, we re-run our analysis of non-SB obscured AGN while including potential starburst galaxies in the obscured sample. In Section \ref{sec:rmsbs}, we identified 10 starbursts in the HST-covered sub-sample of the L20 obscured AGN sample. We input these additional 10 obscured AGN to the non-SB obscured AGN sample used in this study, while also including 10 (non-starbursting) redshift, and F160W matched to the control galaxy sample. We do to this to mimic the effect of studies that do not take into account the presences of SB in their sample creation, and may have uneven amounts of SBs between their science sample and control sample. As seen in \autoref{fig:wstarbursts}, with the inclusion of only 10 star-bursting galaxies, the obscured AGN merger fraction increases by 8\%, and if the results were taken at face value this would imply a 2$\sigma$ excess in the merger fraction of the obscured AGN sample with respect to the control sample. However, this is only due to a bias resulting from the inclusion of the starburst galaxies, and not to any intrinsic physical association between obscured AGNs and mergers. In other words, we are only seeing the possible connection between starbursts and mergers, and no information on the role of mergers in triggering AGNs could be derived by such an analysis. Though it remains to be seen whether the AGN that are triggered by significant mergers are those with SBs in their host galaxies. Due to the lack of sufficient data at hand, we do not compare the merger fractions of obscured AGN with SBs as compared to control galaxies with SBs.          

Instead, as we have shown above, we find that heavily non-SB obscured AGN (mean N$_{H} = 1e24$ erg s$^{-1}$) are not associated in heavily merging systems more than their inactive galaxy counterparts. One major implication of our finding is that the cause of obscuration in most non-SB obscured AGN does not seem to be linked to the funneling of large quantities of gas and dust due to a significant merger as theorized by \citet{hopkins08} and others. AGN may also appear to be obscured due to the orientation of either the torus or the host galaxy itself. Star-forming, inactive galaxies usually are observed being characterized by column densities on the order of $>10^{23}$ cm$^{-2}$ when viewed completely edge on. A notable example of this is the Milky Way. At redshifts higher than this work (i.e z = 4), where galaxies can be extremely dust rich, \citet{circosta2019} measure Compton thick AGN-like obscuration (i.e N$_{H}$  $> 10^{24}$ cm$^{-2}$) in non-AGN galaxies. 

In summary, our results disfavor the major-merger driven non-SB obscured AGN paradigm as the dominant process behind AGN triggering and the cause of the obscuration. As shown in \citet{lambrides20}, the population of obscured AGN in this sample is representative of the lower to moderate luminosity regime of obscured AGN. This regime makes up the predicted bulk of the obscured AGN population as estimated by X-ray background models \citet{gilli07}. The similar merger rates for the obscured sources and the control sample indicate that most obscured AGN are not correlated with major-mergers. Our work does not rule out whether the merger-paradigm works for the highest end of the AGN luminosity or SMBH mass distribution, but other works do \citep[i.e][]{villforth17,marian19}. As previously mentioned, the region of the AGN luminosity parameter space our sample includes represent the bulk of AGN activity at these redshifts.  

It is also possible that minor mergers play a role in triggering AGN. Theoretically, these minor mergers and fly-bys may be able to trigger a disk instability which would ultimately cause the funneling of gas and dust towards the center \citep{hopkinshernquist06,hopkinshernquist09}. At z $>$ 2, simulations find small mergers (M1/M2 $<$ 1/4), are the most frequent \citep[][]{rodriguez2015}. In contrast, \citet{mcalpine2020}, find that galaxy mergers in the EAGLE simulations with mass ratios between 0.1 and 0.25 are not a statistically relevant fueling mechanism for SMBHs. These minor fly-bys and/or mergers are difficult to identify at these redshifts with the data in hand. Future work will entail exploring the fraction of minor mergers in obscured AGN systems, and quantifying the ability for human classifiers to separate minor from major merging systems. Additionally, we plan to carefully analyze the star-formation properties of AGN within and without star-bursting host galaxies in the context of a galaxy's morphology.

\section{Summary and Conclusion}

We test a key prediction of the AGN-Merger paradigm that connects nuclear obscuration of AGN as a consequence of a significant galactic merger. Using a sample of 40 non-starbursting low to moderate luminosity obscured AGN in the GOODS-S field at $0.5 <$z$<3.1$ derived from the deepest X-ray survey to date, we construct a study to test if non-SB obscured AGN are found predominately in major-merging systems. We construct a redshift, magnitude matched inactive galaxy control sample comprised of 40 non-starbursting galaxies. Due to the higher redshifts probed in the sample, we are probing AGN host galaxies that are ill-suited for the most common automated merger identification schemes, and thus we use a sample of 14 expert human classifiers to visually identify the merger status of each galaxy. We estimate each individual classifier's accuracy at identifying merging galaxies/post-merging systems, and isolated galaxies. We calculate the probability of each galaxy being in either a merger or in an isolated system where merger is defined as a galaxy that can either be in a major merger, minor merger, or majorly disturbed system. We do not find any statistically significant evidence that non-SB obscured AGN are predominately found in systems with evidence of significant merging/post-merging features. We further split the sample into different bins of galaxy properties and confirm that is not evidence for statistically significant merger enhancement in non-SB obscured AGN galaxies. 

\acknowledgments
We thank the anonymous referee for their thoughtful insight and important contributions to this work. ELL is supported by STSCI's Director's Discretionary Fund (DDRF) with account number D0101.90261. RG acknowledges support from the agreement ASI-INAF n. 2017-14-H.O. This research has made use of the NASA/IPAC Infrared Science Archive, which is operated by the Jet Propulsion Laboratory, California Institute of Technology, under contract with the National Aeronautics and Space Administration. This publication makes use of data products from the \textit{Wide-field Infrared Survey Explorer}, which is a joint project of the University of California, Los Angeles, and the Jet Propulsion Laboratory/California Institute of Technology, funded by the National Aeronautics and Space Administration. This publication makes use of data products from the Two Micron All Sky Survey, which is a joint project of the University of Massachusetts and the Infrared Processing and Analysis Center/California Institute of Technology, funded by the National Aeronautics and Space Administration and the National Science Foundation. We acknowledge the extensive use of the following Python packages: \software{pandas, scipy, ipython, matplotlib, pymc} \citep[respectively]{pandas, scipy,ipython,matplotlib}. This research made use of \texttt{astropy}, a community-developed core Python package for Astronomy \citep{astropy}.

\appendix
\section{Demo Survey} \label{sec:appendix}

In the following figure, we show the demo survey with answers given to the human classifiers prior to classifying the full sample.  

\begin{figure*}
\centering
\includegraphics[scale=.8,angle=90]{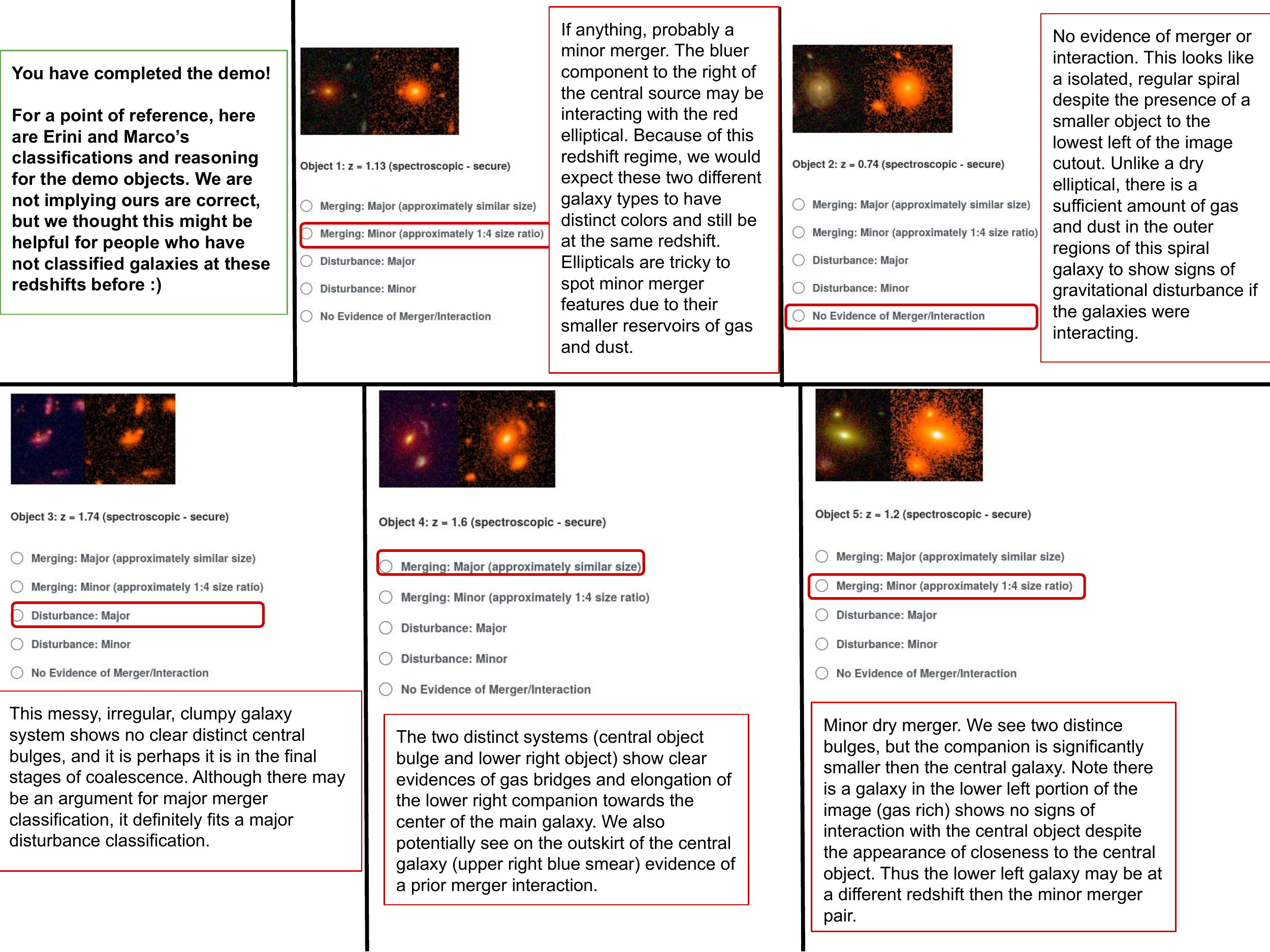}
\label{fig:demo}
\end{figure*} 

\newpage

\bibliography{references}{}
\bibliographystyle{aasjournal}



\end{document}